\documentclass[prd,
 reprint,
superscriptaddress,
preprintnumbers,
nofootinbib,
 amsmath,amssymb,
 aps,
]{revtex4-1}
\usepackage[ocgcolorlinks]{hyperref}
\usepackage{graphicx}

\usepackage[utf8]{inputenc}

\newcommand{\be}{\begin{equation}}
\newcommand{\ee}{\end{equation}}
\newcommand{\ba}{\begin{eqnarray}}
\newcommand{\ea}{\end{eqnarray}}
\renewcommand{\d}{\partial}
\renewcommand{\l}{\left(}
\renewcommand{\r}{\right)}

\newcommand{\e}{\mathrm{e}}

\usepackage[compat=1.1.0]{tikz-feynman}

\begin{document}

\preprint{INR-TH-2020-022}

\title{Probing light exotics from a hidden sector at $c$-$\tau$ factories \\ with polarized electron beams}

\author{Dmitry Gorbunov}
\email{gorby@ms2.inr.ac.ru}
\affiliation{Institute for Nuclear Research of the Russian Academy of Sciences, 117312 Moscow, Russia}
\affiliation{Moscow Institute of Physics and Technology, 141700 Dolgoprudny, Russia}
\author{Dmitry Kalashnikov}
\email{kalashnikov.d@phystech.edu}
\affiliation{Moscow Institute of Physics and Technology, 141700 Dolgoprudny, Russia}
\affiliation{Institute for Nuclear Research of the Russian Academy of Sciences, 117312 Moscow, Russia}

\date{\today}

\begin{abstract}
 Future $c$-$\tau$ factories are natural places to study extensions of
 the Standard Model of particle physics (SM) with new long-lived
 feebly interacting particles light enough to be produced in
 electron-positron collisions. We investigate prospects of these
 machines in exploring such extensions emphasizing the role of
 polarized beams in getting rid of the SM irreducible background for
 the missing energy signature. We illustrate this on the example of $c$-$\tau$ project in Novosibirsk, where the electron beam is designed to be polarized to achieve much higher sensitivity to hadronic resonances and $\tau$-leptons. We investigate models with hidden photons, with millicharged particles (fermions and scalars), with $Z'$ bosons and with axion-like particles. We find that the electron beam polarization of 80\% significantly improves the chances to observe the signal, especially with large statistics. We outline the regions of the model parameter space which can be reached at this factory in one year and in ten years of operation according with the scientific schedule of tuning the energy of colliding beams. 
\end{abstract}

\maketitle

\section{Introduction}
\label{sec:Intro}

Electron-positron colliders are probably the best tools to explore precisely  the Zoo of hadronic resonances, especially if the energy of colliding leptons (the beam energy) may be tuned at will. The latter option is intrinsic to presently operating low-energy $e^+$-$e^-$ machines in Novosibirsk (VEPP-2000, VEPP-4M) and Beijing (BEPC II) and proposed future super $c$-$\tau$ factories in China\,\footnote{For recent presentation see \texttt{https://indico.nucleares.unam.mx} \texttt{/event/1488/session/6/contribution/43}} and Russia\,\footnote{See many details at \texttt{https://sct.inp.nsk.su/} and earlier version at \texttt{https://ctd.inp.nsk.su/c-tau/}}. 

The physical programs of these projects\,\cite{Luo:2018njj,Charm-TauFactory:2013cnj} include thorough investigations of rare processes, which rates within the SM are strongly suppressed. In a realistic case of limited statistics of electron-positron collisions 
these processes give a higher chance to observe new physics. Among them the most encouraging are processes with missing energy, since they may be initiated by a direct production of new light feebly interacting particles  either decaying promptly into invisible mode or long-lived enough to escape from the main detector. The absence of any clear evidences for such pattern in previous experiments may be naturally explained if they are only feebly coupled to the SM particles. We refer to recent reviews\,\cite{Lanfranchi:2020crw,Agrawal:2021dbo,Feng:2022inv} for detailed discussion of relevant models of new physics, searches for the corresponding light hypothetical particles at previous experiments and further prospects with new projects. As to physical motivations, in brief, one may advertise the models with new light hypothetical particles because the SM fails to explain some phenomena like dark matter and neutrino oscillations and yet the quantum corrections from heavy new particles are generally dangerous for the SM Higgs boson mass, see e.g. Refs.\,\cite{Vissani:1997ys,deGouvea:2014xba,Craig:2022uua}.  

To be of practical use in searches for new physics, the missing energy
event must contain at least one observable particle to ensure that
$e^+e^-$ collision happened indeed. The missing energy may be well
explained without any new physics either by production of neutrinos
avoiding any detection or by the detector failure in particle
registration. The first class of events forms a so-called irreducible
background for the searches of new physics events. The second class of
events forms a so-called reducible background, which impact on the searches can be reduced by adjusting the phase space region where the interesting events are accounted for the data analysis. The lower is the total background, the better are our chances to observe new physics events and either measure or constrain parameters of the models with new light particles.      

The colliding leptons may be polarized, which is actually a favorite option of the future projects, see e.g.\,\cite{List:2020wns}. In particular, the Super Charm-Tau  Factory\,\cite{Charm-TauFactory:2013cnj} (hereafter we call it SCTF) plans to operate with 80\% polarization of the electron beam, see e.g.\,\cite{Bondar:2019zgm,Barniakov:2019zhx}. The polarization helps to increase the factory's potential in investigating the Zoo of hadronic resonances in 4-7\,GeV energy range, measuring the weak mixing angle and exploring the $\tau$-lepton physics. Moreover, the direction of polarization (along the beam line) can be reversed at will for any bunch of accelerated electrons, which allows one to improve the particle recognition and keep under control subtle and evasive background processes. 

Remarkably, the polarization can also enhance the factory's sensitivity to light hypothetical particles via suppression of the irreducible background, see e.g.\,\cite{Blaising:2021vhh,Ma:2022cto}. Here we illustrate this point on the example of missing energy events
\begin{equation}
\label{missing}
    e^+\,e^- \to \gamma + \text{missing}\, E\,.
\end{equation}
In case of the unpolarized beams at a $c$-$\tau$ factory this signature has been previously exploited for SM extensions with hidden photons\,\cite{Zhang:2019wnz} and  millicharged fermions\,\cite{Liu:2018jdi,Liang:2019zkb} (see also \cite{Liu:2019ogn}).  
In this paper we estimate the sensitivity of a future Super $c$-$\tau$ factory to the physical parameters of these models, models with millicharged scalars, models with $Z'$ vector boson and models with axion-like particle adopting the realistic setup of the BINP project SCTF \cite{Piminov:2018hgx,Epifanov:2020elk} with 80\% polarized electron beam. 

In all the cases we find significant improvements with respect to the unpolarized operation mode. The reverse option will help to distinguish between the types of new particles as soon as any signal \,\eqref{missing} is  observed. 
Also we repeat our study for the unpolarized beam to check the
previous results (if they exist) and demonstrate the advantage of using the polarized beams especially with large statistics of collisions. Finally, to illustrate the prospects of SCTF in probing these models we take into account the realistic schedule of the machine operation. Indeed the project's main goal is to explore the hadronic resonances and thus the beam energy will be tuned to the corresponding thresholds. At each position a considerable statistics must be collected before shifting to other thresholds. Since the cross sections of both signal and background processes are energy-dependent, the prospects of searches for new physics at the factory depend on its work schedule.  

The paper is organized as follows. In Sec.\,\ref{sec:Background} we investigate the main processes which trigger the background events for the signal of a single photon and missing energy \eqref{missing}. We introduce the set of optimal cuts to suppress the background and improve the physics performance for the searches of new light particles with missing energy signature \eqref{missing}. Sections\,\ref{sec:Photon}-\ref{sec:Axion} are devoted to models with light hidden photon, millicharged particles (fermions and scalars), $Z$'-boson and axion-like particle, respectively. In each case we describe the model, give some physical motivation and relevant references to the origin of constraints on the model parameters in the interesting regions. Then we calculate the signal cross section, evaluate the number of signal events within the chosen cuts and compare the results with those expected from the background study. Further, we outline the expected at 95\% C.L. exclusion regions in the corresponding model parameter space for the machine operating at thresholds of a set of hadronic resonances. Finally, we show the regions expected to be probed with joint analysis of all the data collected over one and ten years of operation. 
We conclude in Sec.\,\ref{sec:Discussion} and discuss additional 
options which might be useful in searches for new physics exploiting the missing energy signature \eqref{missing}.



\section{Expected background}
\label{sec:Background}

In the SM the colliding electron-positron pair may disappear producing
the neutrino-antineutrino pair via either $s$-channel exchange of $Z$-boson or $t$-channel exchange of $W$-boson. If before disappearing a charged lepton emits a photon, 
\begin{equation}
\label{irr-back} 
\e^+\,e^-\to \gamma\,\nu\,\bar\nu\,,
\end{equation}
it gives exactly the event mimicking what we are looking for \eqref{missing}, and hence forms the irreducible background to the search for light exotics. Fortunately, the neutrino production involves only weak interactions, and since (anti)neutrinos are (right)left-handed fermions, the chirality of initial lepton states can be adjusted to suppress the production rate of \eqref{irr-back}. 

In our study we do not tune the lepton polarizations, but rather fix them in accordance with plans of SCTF, where the positron beam is unpolarized, while the electron beam polarization is 80\%.  The polarized electron spin is along the beam axis (the longitudinal polarization) and polarization of any electron bunch can be made either right-handed, positive, (the spin and 3-momenta directions coincide) or left-handed, negative, at will. We observe that the neutrino production \eqref{irr-back} gets suppressed to the level negligible for the light exotic searches. 

The main source of the background for the monophoton and missing energy signature \eqref{missing} of light exotics is associated with non-ideality of the detector. Events with three SM particles in the final state like purely electromagnetic 
\begin{equation}
    \label{red-back}
 e^+ e^- \rightarrow \gamma l^+l^-\,, \;\;\;    e^+ e^- \rightarrow \gamma\gamma\gamma 
\end{equation}
easily mimic the signal signature \eqref{missing}, if the lepton (electron and muon) pair or a couple of photons escapes detection. Indeed, apart from possible detector malfunctioning, there is always a chance that the lepton (photon) pair is produced in a specific part of the phase space with 3-momenta almost along the beam axis, where its detection is technically impossible.  

\begin{figure}[!htb]
    \centerline{
    \includegraphics[width=1.4\columnwidth]{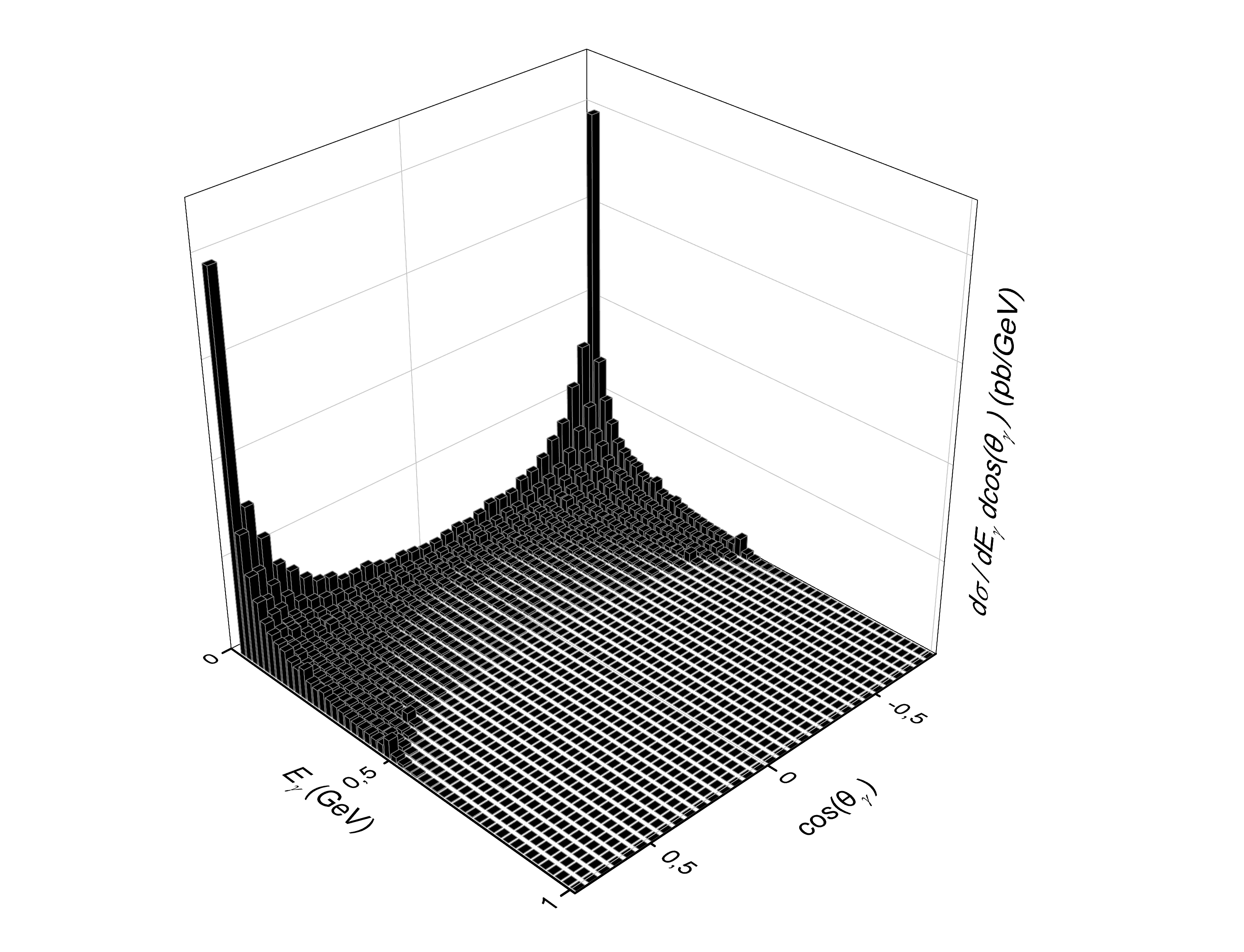}}  \caption{Distribution of photons over angle $\theta_\gamma$ and energy $E_\gamma$ in $e^+e^-\to \gamma e^+e^-$ with collision energy $\sqrt{s}=3$\,GeV and lepton angles obeying $|\cos\theta_l|>0.95$.}
    \label{fig:2dim}
\end{figure}

\begin{figure}[!htb]
    \centerline{
    \includegraphics[width=1.1\columnwidth]{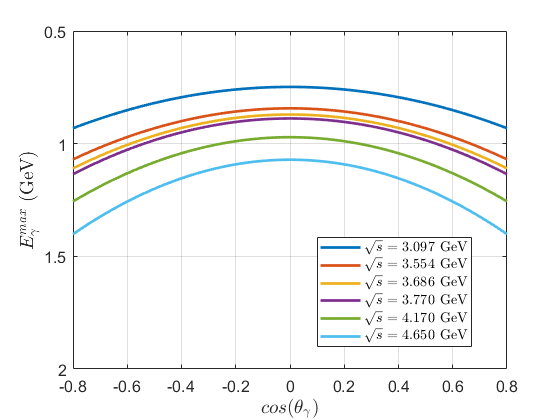}}
    \caption{Maximal photon energy $E_\gamma^{\text{max}}$ as a function of photon angle $\theta_\gamma$.}
    \label{fig:Emin}
\end{figure}

Events of the type \eqref{red-back} form the reducible background,
which impact on the light exotic searches can be diminished to some
extent by choosing the appropriate cuts on the energy and direction of
the single photon of signal event \eqref{missing}. To this end we use
CalcHEP package\,\cite{Belyaev:2012qa} and simulate a set of
collisions with a photon and an electron-positron (muon-antimuon) pair
in the final state \eqref{red-back}. We observe the muon contribution
to be  negligible as compared to that of the electron. We suppose that a produced lepton with 3-momentum forming an  angle $\theta$ with the beam direction can be detected and will not be missed if $|\cos\theta_l|<0.95$, the distribution of photons over angle and energy is depicted in Fig.\,\ref{fig:2dim}. 
Considering such events we find that for each angle $\theta_\gamma$ between the photon and beam axis the photon energy is limited from above by a certain value which depends on the angle, $E^{\text{max}}_\gamma(\theta_\gamma)$.  There are simply no events with more energetic photons. In what follows we constrain ourselves to the phase space region with photons obeying 
\begin{equation}
\label{photon-cut}
|\cos\theta_\gamma|<0.8 \,,      
\end{equation}
with maximal energy inferred from the simulations we numerically approximate as a function of angle $\theta_\gamma$ and present in Fig.\,\ref{fig:Emin}.  
We checked that the same kinematic bounds reduce the background events from the second process in \eqref{red-back} as well. The maximal energy scales linearly with the collision energy, $E^\text{max}_\gamma\propto\sqrt{s}$. The relevant kinematics of the electromagnetic processes \eqref{red-back} does not depend on the beam polarization.

Therefore, below, in our investigation of the signal events \eqref{missing} expected in a set of SM extensions we limit the photon angle as \eqref{photon-cut} and photon energy as 
\begin{equation}
    \label{photon-energy}
    E_\gamma(\theta_\gamma)>E_\gamma^\text{max}(\theta_\gamma)\,.
\end{equation}

When the missing energy in the signal process \eqref{missing} implies
production of a single hypothetical particle, one can additionally
suppress the background from neutrino production \eqref{irr-back}
using the fact that for processes with 2 particles in the final state the energy of signal photon $E_\gamma$ is actually fixed by kinematics. It equals 
\begin{equation}
\label{Egamma-2-2}
 E_\gamma = \frac{s - m^2}{2\sqrt{s}}\,,   
\end{equation}
where $m$ is mass of the hypothetical particle. In practice, the
electromagnetic calorimeter employed to determine the photon energy
has a finite energy resolution, which in the case of SCTF is\,\cite{Aihara:2019lzh} 
\begin{equation}
\label{add-cuts}
\begin{split}
& \sigma(E_\gamma) = E_\gamma\times \l 0.019\times\l \frac{1 \, \text{GeV}}{E_\gamma}\r^{1/4} \right. \\ 
& \left. +0.0033\times \l \frac{1 \, \text{GeV}}{E_\gamma}\right)^{1/2}+0.0011 \times\l\frac{1 \, \text{GeV}}{E_\gamma}\r \r. 
\end{split}
\end{equation}
Therefore, to evaluate the background events we average the final photon energy with Gaussian distribution defined by the dispersion \eqref{add-cuts} and mean value \eqref{Egamma-2-2}. It enables us to greatly reduce the number of background events for the $2\to2$ signal processes.  

Then, to estimate the sensitivity of the future $c$-$\tau$ factory
SCTF to the models with light exotic particles we adopt its planned
operation schedule\,\cite{Epifanov:2020elk}, which implies collecting
a certain amount of $e^+$-$e^-$ collisions at a given energy and then shifting to other energies. The collision energies are close to the thresholds of interesting resonances, see Tab.\,\ref{Tab:schedule}. 
\begin{table}[!htb]
    \centering
    \begin{tabular}{|c|c|c|c|c|c|c|}
    \hline
    $\sqrt{s}$, \, GeV & $3.097$ & $3.554$ & $3.686$ & $3.770$ & $4.170$ & $4.650$ \\
    \hline
    $L, \, \text{fb}^{-1}$ & $300$ & $50$ & $150$ & $300$ & $100$ & $100$ \\
    \hline
    \end{tabular}
    \caption{Annual operation schedule of SCTF: sets of the collision energy $\sqrt{s}$ and the corresponding integrated luminosity $L$.}
    \label{Tab:schedule}
\end{table}
We find that with our chosen cuts \eqref{photon-cut} and \eqref{photon-energy}, each of the six operation stages of Tab.\,\ref{Tab:schedule} are background-free for our study: there are no contributions from both reducible \eqref{red-back} and irreducible \eqref{irr-back} backgrounds for the statistics of electron-positron collisions expected to be collected at SCTF with 80\% polarized electrons, see Tab.\,\ref{Tab:BG}. Hence, to estimate the sensitivity of the $c$-$\tau$ factory to the new light particles in each model to be achieved at each of the operational stages, below we calculate the production of the single photon and corresponding light particles in electron-positron scattering and tune the model parameters to obtain 3 signal events \eqref{missing}. Within the Poisson statistics it implies that models with larger coupling constants will be excluded at 95\% C.L. if no events are observed. 

Moreover, after one year of operation in accordance with the plan of Tab.\,\ref{Tab:schedule}, we can place stronger bounds on the models with light exotics  
performing a joint analysis of data from all six stages. The operation stages are independent, and so for each model we simply sum up the events from all the stages. However, the number of expected background events exceeds one in this analysis. We make use of the Feldman--Cousins approach \cite{Feldman:1997qc,Hill:2002nv} to assess the prospects in testing the new models. In case of {\it observed} $n$ events with expected $B$ events from the background, the number of signal events $S$ can be constrained at the $\alpha$\,C.L. from the Poisson distribution as  
\begin{equation}
    \label{P}
    1-\alpha=\frac{\e^{-B-S}\sum_{i=0}^n\frac{(B+S)^i}{i!}}{\e^{-B}\sum_{i=0}^n\frac{B^i}{i!}}\,.
\end{equation}
Note that with $B=0$ eq.\,\eqref{P} gives 3 for no events, $n=0$, and $\alpha=0.95$, as we use to get the bounds at 95\%\,C.L. in the background-free case. Therefore, to evaluate the expected 95\%\,C.L.-limit on the number of signal events to be obtained with small but non-zero background, $B\gtrsim 1$, for several numbers of observed events $n$ we calculate the value of signal $S=S(n)$ from eq.\,\eqref{P} and then average them with the Poisson distribution for $n$ at zero signal as follows 
\begin{equation}
    \label{AV}
    \bar S= \e^{-B} \sum _{n=0}^\infty \frac{S(n)B^n}{n!}\,.
\end{equation}
So, for each of the models we tune the model parameters to get the number of signal events equal $\bar S$: the models with stronger couplings will be excluded at 95\%\,C.L. after one year of operation of SCFT following the schedule of Tab.\,\ref{Tab:schedule}. 

Finally, we estimate the reach of SCFT after 10 years of the operation with total integrated luminosity  of 10\,ab$^{-1}$. To this end we sum separately signal and background events following the annual plan of Tab.\,\ref{Tab:schedule} and apply eqs.\,\eqref{P}, \eqref{AV} in the models where the number of expected background events does not exceed ten. These are all the cases where the signal processes are $2\to2$. In the models with more particles in the final states the background is higher, the number of expected background events $B$ significantly exceeds ten, 
which one can realize from the estimates of the background events for one year of  operation shown in Tab.\,\ref{Tab:BG}. Then we accept the Gaussian statistics, tune the parameters of the models with light hypothetical particles to obtain $S$ signal events, such that 
\[
\frac{S}{\sqrt{B}}=2\,,
\]
and say that all the models with larger couplings will be ruled out at 95\%\,C.L. 

\begin{table}[!htb]
    \centering
    \begin{tabular}{|c|c|c|c|c|c|c|}
    \hline
    $\sqrt{s}$, \, GeV & $3.097$ & $3.554$ & $3.686$ & $3.770$ & $4.170$ & $4.650$ \\
    \hline
    2 particle f.s. & $0.097$ & $0.02$ & $0.066$ & $0.14$ & $0.054$ & $0.066$ \\
    \hline
    3 particle f.s. & $2.8$ & $0.63$ & $2.1$ & $4.4$ & $1.8$ & $2.3$ \\
    \hline
    \end{tabular}
    \caption{Approximate values of irreducible background events for the models with mass of the hidden particle $m=100$\,MeV and signal events with 2 or 3 particles in the final state (f.s.). We assume 80\% longitudinal polarization of the electron beam and the collision statistics for one year of operation in accordance with the annual operation schedule of SCTF, see Tab.\,\ref{Tab:schedule}.}
    \label{Tab:BG}
\end{table}

\section{Hidden Photon}
\label{sec:Photon}

The new physics we are looking for may  be located in a hidden sector, which is not connected with known physics, that is the SM sector, via known SM gauge interactions. However, some connection apart of via gravity still may exist, and for light feebly interacting particles from the hidden sector the most promising (as concerns direct searches) are those provided by {\it portals}, which are renormalizable dimension four interaction terms involving fields from the both sectors. In particular, so-called {\it vector portal} describes interaction between the SM hypercharge vector field $A_\mu$ and a hidden vector field $A_\mu'$, \cite{Holdom:1985ag}
\begin{equation*}
    {\cal L}_{\text{int}}\propto \varepsilon \times\l \d_\mu A_\nu-\d_\nu A_\mu\r 
 \l \d_\mu A'_\nu-\d_\nu A'_\mu\r,    
\end{equation*}
with dimensionless parameter $\varepsilon\ll 1$. It mixes kinetic terms of the two vectors and is invariant with respect to Abelian gauge transformations in both sectors, $A_\mu^{(')}\to A_\mu^{(')}+\d_\mu\alpha^{(')}$. 

At low energies relevant for $c$-$\tau$ factories this mixing yields (upon field redefinition, see e.g.\,\cite{Redondo:2008aa}) an effective interaction between the new light vector and the SM electromagnetic current.     
The interesting coupling is that to electrons and positrons, 
\begin{equation}
\label{Aee}
    \mathcal{L} = \varepsilon e A^{'}_\mu \Bar{e} \gamma^\mu e\,,
\end{equation}
where $e$ is the electric charge of the positron. This coupling induces production of $A_\mu'$ in $e^+$-$e^-$ scattering. Therefore, if $\varepsilon\ll1$ and $A'$ is either stable or decays mostly into invisible particles from the hidden sector, one can suggest the signal \eqref{missing} as a signature of this light vector. 

In our study we adopt a minimal phenomenological model, where the light vector is characterized by its mass $m_{A'}$ and interaction \eqref{Aee}, and we neglect the  induced by \eqref{Aee} decay of $A'$ into SM fermions. In literature a light vector with tiny coupling is usually dubbed as {\it paraphoton}, {\it hidden} or {\it dark photon} \cite{Okun:1982xi}. It is predicted in SM extensions and may be related to some unexplained phenomena, including the dark matter \cite{Pospelov:2007mp}, see Refs.\,\cite{Graham:2021ggy,Filippi:2020kii} for recent reviews. More complicated signatures than \eqref{missing} can be advertised in particular models, see e.g. Ref.\,\cite{Boos:2022gtt} for a recent investigation of SCTF prospects.

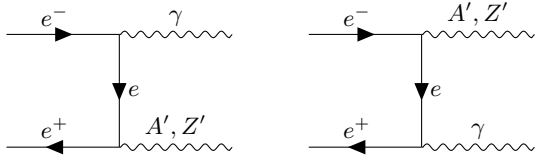
\begin{figure}[!htb]
    \centering
    \begin{tikzpicture}
\begin{feynman}
    \vertex at (0,0) (i1);
    \vertex at (0,-1.5) (i2);
    \vertex at (1.5,0) (a1);
    \vertex at (1.5,-1.5) (a2);
    \vertex at (3,-1.5) (f2);
    \vertex at (3,0) (f1);
    
    \diagram*{
    (i1) -- [fermion, edge label={\!\!\!\!\(e^{-}\)}] (a1),
    (i2) -- [anti fermion, edge label={\!\!\!\!\(e^{+}\)}] (a2),
    (a1) -- [fermion, edge label={\(e\)}] (a2),
    (a1) -- [photon, edge label={\(\gamma\)}] (f1),
    (a2) -- [photon, edge label={\(A^{\prime}, Z^{\prime}\)}] (f2),
    };
\end{feynman}
\end{tikzpicture}
\hskip 0.05\textwidth 
\begin{tikzpicture}
\begin{feynman}
    \vertex at (0,0) (i1);
    \vertex at (0,-1.5) (i2);
    \vertex at (1.5,0) (a1);
    \vertex at (1.5,-1.5) (a2);
    \vertex at (3,-1.5) (f2);
    \vertex at (3,0) (f1);
    
    \diagram*{
    (i1) -- [fermion, edge label={\!\!\!\!\(e^{-}\)}] (a1),
    (i2) -- [anti fermion, edge label={\!\!\!\!\(e^{+}\)}] (a2),
    (a1) -- [fermion, edge label={\(e\)}] (a2),
    (a1) -- [photon, edge label={\(A^{\prime}, Z^{\prime}\)}] (f1),
    (a2) -- [photon, edge label={\(\gamma\)}] (f2),
    };
\end{feynman}
\end{tikzpicture}

    \caption{The Feynman diagrams for the production of dark photon $A'$ or $Z^\prime$ boson and a photon.}
    \label{diagram:dp}
\end{figure}

One can examine the production of a hidden photon associated with a
single photon. It is the $2\to2$ process, see Fig.\ref{diagram:dp}, 
and at each direction of the outgoing photon parametrized by the polar angle $\theta_\gamma$ the photon energy is fixed by the angle and the collision energy as \eqref{Egamma-2-2}. The corresponding differential cross section in the limit of zero electron mass reads\,\cite{Zhang:2019wnz}
\begin{equation}
\label{cross-A}
    \frac{d\sigma}{d\cos\theta_\gamma} = \frac{2\pi \varepsilon^2 \alpha^2}{s} \left( 1 - \frac{m_{A'}^2}{s} \right) \frac{1+\cos^2\theta_\gamma+\frac{4sm_{A'}^2}{(s-m_{A'}^2)^2}}{(1-\cos^2\theta_\gamma)}\,.
\end{equation}
Expectedly, the cross section of this pure electromagnetic process does not depend on the polarization of initial leptons. 

\begin{figure}[!htb]
    \centerline{
    \includegraphics[width=1.1\columnwidth]{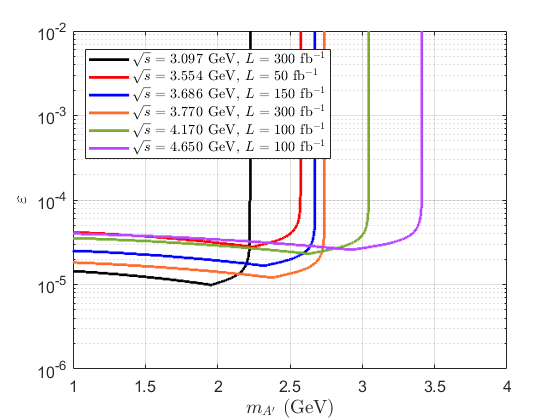}}
    \caption{The regions above the lines can be tested at 95\%\,C.L. with $e^+ e^- \rightarrow \gamma A^{'}$.}
    \label{fig:DarkPhoton}
\end{figure}

Imposing the cut on photon angle\,\eqref{photon-cut} we integrate \eqref{cross-A} over $\theta_\gamma$ and obtain the number of signal events. Then, by setting it equal 3 we get the 95\% C.L. exclusion limits outlined in Fig.\,\ref{fig:DarkPhoton} 
for a set of collision energies and integrated luminosities consistent with the working schedule of SCTF presented in Tab.\,\ref{Tab:schedule}. 
Note that the limits in Fig.\,\ref{fig:DarkPhoton} become stronger
with growing mass, which is anticipated from the term with denominator
$(s-m_{A'}^2)$ in \eqref{cross-A}, and corresponds to the almost
resonant production, $e^+e^-\to A'$, near the threshold. Technically,
it is originated from the singularity in the $t$-channel fermion
propagator. However, our cuts \eqref{photon-cut} and \eqref{photon-energy} exclude that kinematic region from the analysis, and so the maximum sensitivity is achieved at some mass near but not at the threshold. 

\begin{figure}[!htb]
    \centerline{
    \includegraphics[width=1.1\columnwidth]{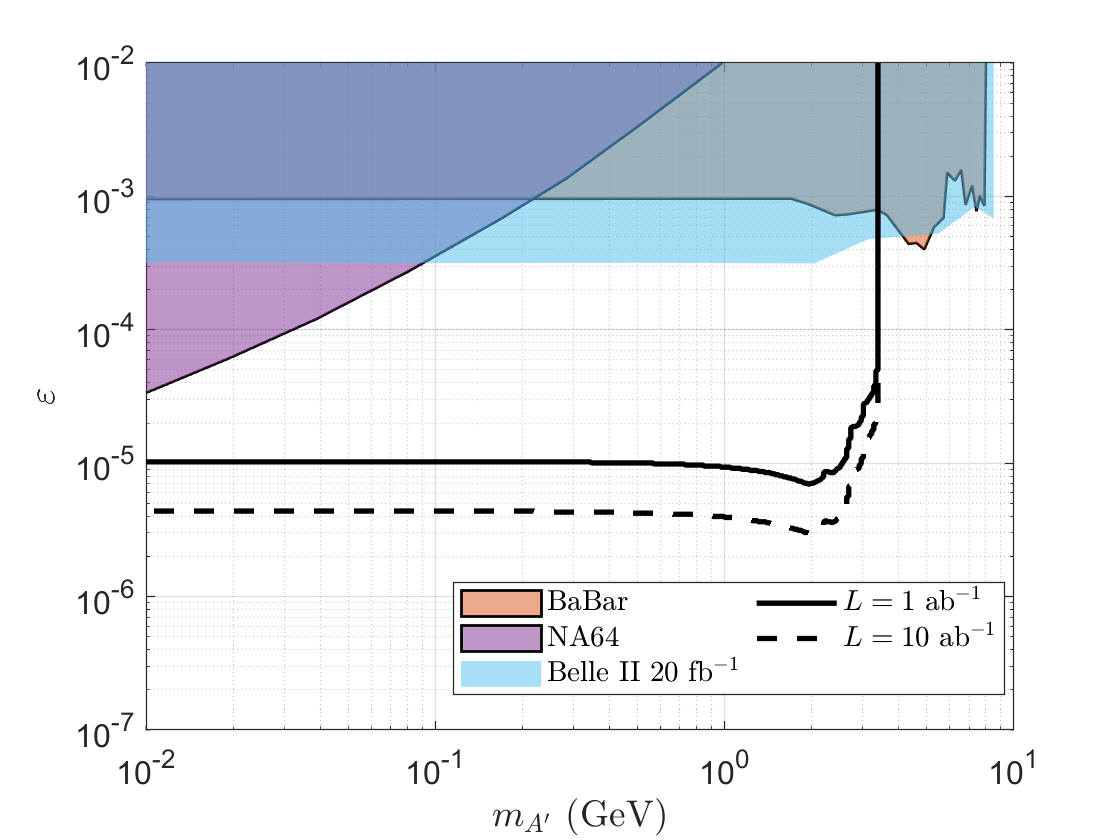}}
    \caption{The regions above the black bold and dashed lines will be
      probed at 95\%\,C.L. with $e^+ e^- \rightarrow \gamma A'$ after
      one and ten years of operation, respectively. The existing
      limits (colored and outlined) and expected reaches of the ongoing Belle-II experiment (colored) are taken from Ref.\,\cite{Kou_2019}.}
    \label{fig:DarkPhoton_sum}
\end{figure}

\begin{figure}[!htb]
    \centerline{
    \includegraphics[width=1.1\columnwidth]{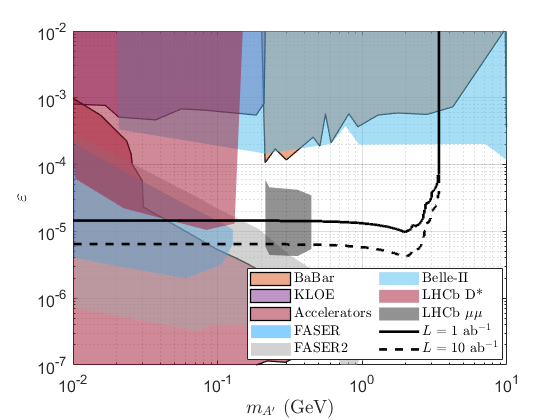}}
    \caption{The regions above the black bold and dashed lines will be probed in the HH case (see the main text for details) at 95\%\,C.L. with $e^+ e^- \rightarrow \gamma A'$ after one and ten years of operation, respectively. The existing limits (colored and outlined) and expected reaches of ongoing experiments (colored) on the visible mode are taken from Ref.\,\cite{Feng:2022inv}.}
    \label{fig:DarkPhoton_sum_vis}
\end{figure}

One can estimate the entire prospects of probing the models with a light hidden photon at SCTF after one and ten years of operation in accordance with the schedule of Tab.\,\ref{Tab:schedule} by making use of the procedure specified in Sec.\,\ref{sec:Background}. The results are depicted in Fig.\,\ref{fig:DarkPhoton_sum} 
together with existing experimental constraints and prospects of ongoing searches. The SCTF can explore a considerable region inaccessible to the ongoing searches. To emphasize the importance of searches for missing energy and monophoton signature \eqref{missing} at SCTF we consider their value in a model where the hidden photon decays into visible modes (e.g. $e^+e^-$) as well, say in 50\% cases. We call it as half-hidden (HH) case in what follows. Then searches for the visible decay modes are relevant as well as searches for the invisible mode. In this model our expected signal with missing energy is 2 times lower, therefore the {\it upper limits} on $\varepsilon$ must be shifted by a factor $\sqrt{2}$. Likewise the {\it lower bounds} on $\varepsilon$ for the visible modes must be shifted by almost the same factor. The corrected in this way constraints from searches for the visible mode and the expected limits from future searches at SCTF for the invisible mode are presented in Fig.\,\ref{fig:DarkPhoton_sum_vis}.     
One concludes that searches at SCTF for the invisible modes are competitive with ongoing searches for the visible modes, if the visible and invisible decay rates of the hidden photon are of the same order. 

Finally, we perform calculations of the signal rate with the same angular and energy cuts and for the same beam energies as were done for a $c$-$\tau$ factory in Ref.\,\cite{Zhang:2019wnz}, where the colliding beams are considered to be unpolarized. The results are in agreement, since the signal does not depend on the polarization. As to the expected limits, they are the same for the small statistics, when the chosen energy and angular cuts are enough to keep the number of expected background events \eqref{red-back} below one for the unpolarized beams. As we explained in Sec.\,\ref{sec:Background}, exploiting polarized beams provides an additional tool to mitigate the background, and so for higher statistics our expected limits are stronger than the similar ones in \cite{Zhang:2019wnz}, especially for ten years of operation.   

\section{Millicharged Particles}
\label{sec:MCP}

Since quantization of the hypercharge of the SM particles may seem
puzzling one can envisage hypothetical particles with arbitrary small
hypercharge \cite{Foot:1990mn}. Consequently, at low energies there
may be particles $\chi$ with arbitrary small electric charge,
generically called {\it millicharged particles}  (MCPs). In the case of fermions, their coupling to SM photon $A_\mu$ is
\begin{equation}
\label{MCPs}
    \mathcal{L} = \varepsilon e A_\mu \Bar{\chi} \gamma^\mu \chi\,,
\end{equation}
where the charge is written in terms of the positron electric charge $e$, and we assume $\varepsilon\ll 1$. It is worth noting that in the hidden sector scenario with vector portal described in Sec.\,\ref{sec:Photon}, any particles from the hidden sector charged under the Abelian group associated with the hidden photon $A'$ become MCPs upon proper redefinition of the vector fields $A_\mu$ and $A_\mu'$.

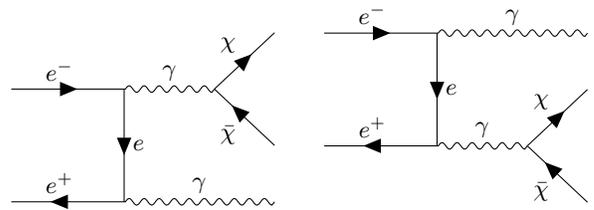
\begin{figure}[!htb]
    \centering
\begin{tikzpicture}
\begin{feynman}
    \vertex at (0,0) (i1);
    \vertex at (0,-1.5) (i2);
    \vertex at (1.5,0) (a1);
    \vertex at (1.5,-1.5) (a2);
    \vertex at (2.7,0) (a3);
    \vertex at (3.5,-1.5) (f1);
    \vertex at (3.5,0.75) (f2);
    \vertex at (3.5,-0.75) (f3);
    
    \diagram*{
    (i1) -- [fermion, edge label={\!\!\!\!\(e^{-}\)}] (a1),
    (i2) -- [anti fermion, edge label={\!\!\!\!\(e^{+}\)}] (a2),
    (a1) -- [fermion, edge label={\(e\)}] (a2),
    (a2) -- [photon, edge label={\(\gamma\)}] (f1),
    (a1) -- [photon, edge label={\(\gamma\)}] (a3),
    (a3) -- [fermion, edge label={\(\chi\)}] (f2),
    (a3) -- [anti fermion, edge label'={\(\bar{\chi}\)}] (f3),
    };
\end{feynman}
\end{tikzpicture}
\hskip 0.03\textwidth 
\begin{tikzpicture}
\begin{feynman}
    \vertex at (0,0) (o);
    \vertex at (0,-0.75) (i1);
    \vertex at (0,-2.25) (i2);
    \vertex at (1.5,-0.75) (a1);
    \vertex at (1.5,-2.25) (a2);
    \vertex at (2.7,-2.25) (a3);
    \vertex at (3.5,-0.75) (f1);
    \vertex at (3.5,-1.5) (f2);
    \vertex at (3.5,-3.0) (f3);
    
    \diagram*{
    (i1) -- [fermion, edge label={\!\!\!\!\(e^{-}\)}] (a1),
    (i2) -- [anti fermion, edge label={\!\!\!\!\(e^{+}\)}] (a2),
    (a1) -- [fermion, edge label={\(e\)}] (a2),
    (a1) -- [photon, edge label={\(\gamma\)}] (f1),
    (a2) -- [photon, edge label={\(\gamma\)}] (a3),
    (a3) -- [fermion, edge label={\(\chi\)}] (f2),
    (a3) -- [anti fermion, edge label'={\(\bar{\chi}\)}] (f3),
    };
\end{feynman}
\end{tikzpicture}

\caption{The Feynman diagrams for the production of $\chi \bar{\chi}$ and a photon.}
 \label{diagram:mcp}
\end{figure}

\begin{figure}[!htb]
    \centerline{
    \includegraphics[width=1.1\columnwidth]{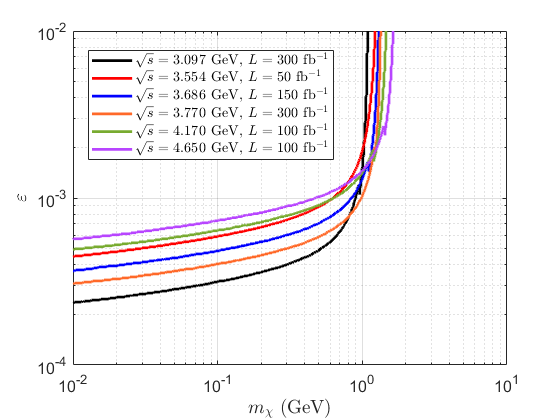}}
    \caption{The regions above the lines will be probed at 95\%\,C.L. with $e^+ e^- \rightarrow \gamma \Bar{\chi} \chi$ in the particular operational modes.}
    \label{fig:Millicharge}
\end{figure}

\begin{figure}[!htb]
    \centerline{
    \includegraphics[width=1.1\columnwidth]{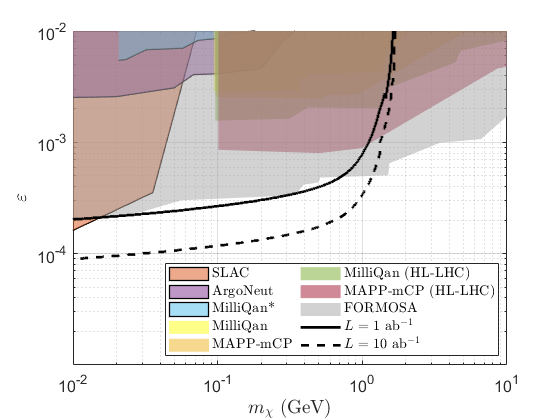}}
    \caption{The regions above the black bold and dashed lines will be excluded at 95\%\,C.L. with $e^+ e^- \rightarrow \gamma \Bar{\chi} \chi$ after one and ten years of operation. The existing limits (colored and outlined) are taken from Ref.\,\cite{Feng:2022inv}, prospects of MAPP-mCP and MilliQan (at the present stage of LHC operation, i.e. Run-3) are from Ref.\,\cite{Acharya:2022nik}, their prospects at LHC operation in high luminosity mode (HL-LHC) and prospects of FORMOSA are copied from Ref.\,\cite{Feng:2022inv}.}
    \label{fig:Millicharge_sum}
\end{figure}

Interaction \eqref{MCPs} gives rise to the MCP production associated
with the photon, see the corresponding Feynman diagrams in Fig.\ref{diagram:mcp}. 
The outgoing MCPs with small charge cannot be observed (see however Ref.\,\cite{Gorbunov:2022bzi} for the special case of very slow MCPs), and so one can use the signature \eqref{missing} to explore the models with MCPs. The signal process is $2\to3$, and hence both energy and angle of the measured photons are independent parameters. The corresponding differential production cross section can be written as\,\cite{Liu:2018jdi}
\begin{equation}
\label{MCP-diff}
\begin{split}
        \frac{d\sigma}{dE_\gamma d\cos\theta_\gamma} = & \frac{8\varepsilon^2\alpha^3 (1+2m_\chi^2/s_\gamma) \beta_\chi}{3sE_\gamma(1-\cos^2\theta_\gamma)} \\
        & \times \left[ 1+\frac{E_\gamma^2}{s_\gamma}(1+\cos^2\theta_\gamma) \right],
\end{split}
\end{equation}
where $s_\gamma \equiv s-2\sqrt{s}E_\gamma$, $\beta_\chi \equiv \sqrt{1-4m_\chi^2/s_\gamma}$. 
Integrating it over the photon energy and angle within the chosen cuts
\eqref{photon-cut} and \eqref{photon-energy}, one arrives at the expected at 95\%\,C.L. limits presented in Fig.\,\ref{fig:Millicharge}. 
The described in Sec.\,\ref{sec:Background} combined analysis of all the data collected at the operation stages from Tab.\,\,\ref{Tab:schedule} yields 
limits expected to be obtained (if no signal events are observed) after one and ten years of operation of SCTF. We outline them in Fig.\,\ref{fig:Millicharge_sum} 
along with established experimental bounds and anticipated reaches of ongoing searches.  

To check our calculations we reproduced the results of Ref.\,\cite{Liang:2019zkb} when placing the same cuts and inserting the same beam energy. There the lepton beams were treated as unpolarized, and so the background was non-negligible even for one year of operation (the collected statistics of about 1\,ab$^{-1}$), though an additional cut has been utilized to suppress it. Thus, we observe that the SCTF (where the electron beam is polarized) will place stronger limits after one and after ten years of operation. An additional cut and a special optimization procedure have been advertised in Ref.\,\cite{Liang:2019zkb} to improve the situation in testing the models with light MCPs. Likewise, it well may increase the sensitivity in case of the longitudinally polarized electron beam of SCTF.   

It is worth noting that with three particles in the final state the 
limiting curves in Figs.\,\ref{fig:Millicharge} and \ref{fig:Millicharge_sum} rise up noticeably with MCP mass. The models with MCP masses close to the threshold, $m_\chi=\sqrt{s}/2$, can be probed by tracing the nonrelativistic MCPs along the lines of Ref.\,\cite{Gorbunov:2022bzi}. 

\begin{figure}[!htb]
    \centerline{
    \includegraphics[width=1.1\columnwidth]{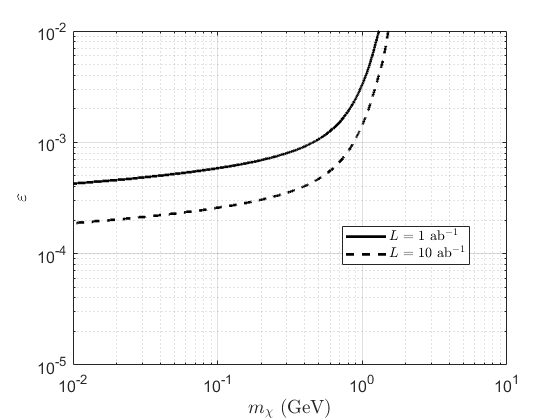}}
    \caption{The regions above the black bold and dashed lines will be excluded at 95\%\,C.L. with $e^+ e^- \rightarrow \gamma \Bar{\chi} \chi$ with scalar mCP after one and ten years of operation.}
    \label{fig:Scalar_Millicharge_sum}
\end{figure}

To complete the task we consider the models with scalars as particles carrying a small electric charge. We obtain for the differential cross section of the scalar MCPs production associated with a single photon the following expression, 
\begin{equation}
\begin{split}
        \frac{d\sigma}{dE_\gamma d\cos\theta_\gamma} = & \frac{2\varepsilon^2\alpha^3 \beta_\chi^3}{3sE_\gamma(1-\cos^2\theta_\gamma)} \\
        & \times \left[ 1+\frac{E_\gamma^2}{s_\gamma}(1+\cos^2\theta_\gamma) \right],
\end{split}
\end{equation}
where we use the same notations as in eq.\,\eqref{MCP-diff}. In the limit of small MCP masses the results differ by the number of final states, that is four. In Fig.\,\ref{fig:Scalar_Millicharge_sum} 
we depict the limits on $\epsilon$ expected to be reached after one and ten years of operation in accordance with the SCTF annual program of Tab.\,\ref{Tab:schedule}.  
\section{$Z^{'}$ boson}
\label{sec:Z'}

There are SM extensions with a light exotic vector, which couples differently to right-handed and left-handed SM fermions, see e.g. 
Refs.\,\cite{He:1991qd,Bodas:2021fsy,Anastasopoulos:2022ywj}.  This
hypothetical particle is generically named as the $Z'$ boson and its interaction with charged leptons can be parametrized with vector $g_V$ and axial $g_A$ couplings as   
\begin{equation}
\label{Z'-coupling}
    \mathcal{L} = Z^{'}_\mu \Bar{e}\gamma^\mu (g_V - g_A \gamma_5) e\,.
\end{equation}
Then lepton scattering may produce $Z'$ boson accompanied by a photon,
which signature is \eqref{missing} if the $Z'$ boson is stable or decays mostly in an invisible mode. Since the strength of $Z'$ coupling to SM fermions \eqref{Z'-coupling} depends on their chirality, so the production cross section does. With the unpolarized positron beam and electron beam polarization $\epsilon \in [-1;1]$ one evaluates the differential cross section $Z'$ as
\begin{equation}
\label{Z-cross}
\begin{split}
    \frac{d\sigma}{d\cos\theta_\gamma} = & \frac{\alpha}{2 s} \times (g_A^2 + g_V^2 - 2 \, \epsilon \, g_A g_V) \\
    & \times \left( 1 - \frac{m_{Z'}^2}{s} \right) \frac{1+\cos^2\theta_\gamma+\frac{4sm_{Z'}^2}{(s-m_{Z'}^2)^2}}{(1-\cos^2\theta_\gamma)}\,.
\end{split}
\end{equation}
It is $2\to2$ scattering, see Fig.\ref{diagram:dp}, hence the photon energy $E_\gamma$ is fixed by the collision energy at any angle $\theta_\gamma$ as \eqref{Egamma-2-2}.   
Formula \eqref{Z-cross} transforms to \eqref{cross-A} with unpolarized
beams, $\epsilon=0$, and also in the limit of pure vector-like
coupling, $g_A\to 0$, where any dependence  on the beam polarization
disappears. The same is true for pure axial coupling. Then, if $Z'$
couples only to left-handed leptons like the SM $Z$-boson, we have
$g_A=g_V$ and the cross section vanishes for the fully positively polarized electron beam, $\epsilon\to 1$. This illustrates our preference for the beam polarization, which guarantees the suppression of the irreducible background expected from the neutrino production \eqref{irr-back}. 

\begin{figure}[!htb]
    \centerline{
    \includegraphics[width=1.1\columnwidth]{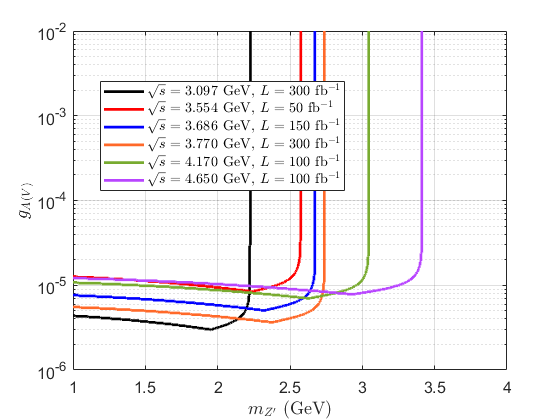}}
    \caption{The parameter regions for models with $g_V = 0$ or $g_A = 0$ 
    to be probed at 95\%\,C.L. with $e^+ e^- \rightarrow \gamma Z^{'}$ and any polarization $\epsilon$ of the electron beam. The limits match those for the hidden photon presented in Fig.\,\ref{fig:DarkPhoton} upon replacement $g=\varepsilon e$. Recall that the irreducible background \eqref{irr-back} gets suppressed with positive polarization,  $\epsilon>0$.}
    \label{fig:ZBoson_gV=gA_H-}
\end{figure}

\begin{figure}[!htb]
    \centerline{
    \includegraphics[width=1.1\columnwidth]{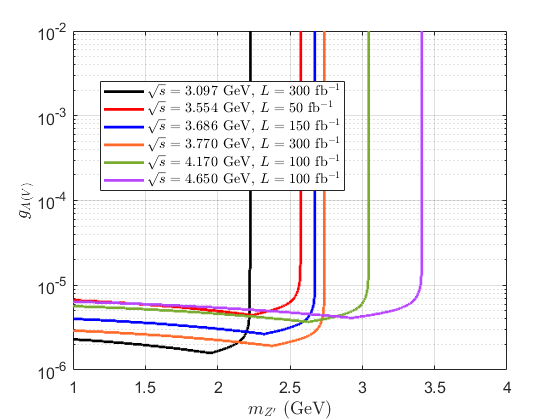}}
    \caption{The regions to be excluded at 95\%\,C.L. with $e^+ e^- \rightarrow \gamma Z^{'}$ with $\epsilon=0.8$ and $g_A=-g_V$, the right-handed type coupling.}
    \label{fig:ZBoson_gA_H-}
\end{figure}

\begin{figure}[!htb]
    \centerline{
    \includegraphics[width=1.1\columnwidth]{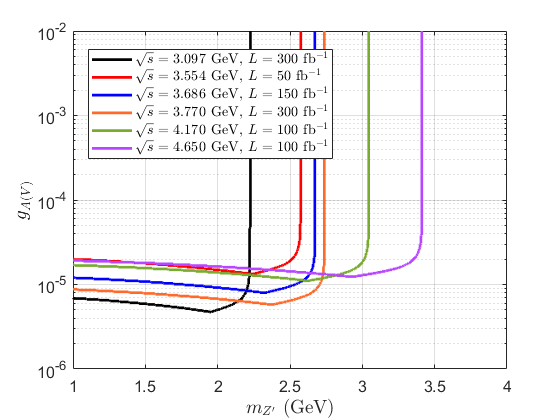}}
    \caption{The regions to be tested at 95\%\,C.L. with $e^+ e^-
      \rightarrow \gamma Z^{'}$ for $\epsilon=0.8$ and $g_V=g_A$, the
      left-handed type coupling. One observes a lower discovery potential as compared to the right-handed type in Fig.\,\ref{fig:ZBoson_gA_H-}.}
    \label{fig:ZBoson_gV-gA_H-}
\end{figure}

Integrating the differential cross section \eqref{Z-cross} within the
angular cuts \eqref{photon-cut} we evaluate the reach of the $c$-$\tau$ factory to the $Z'$ boson couplings, which we show in 
Figs.\,\ref{fig:ZBoson_gV=gA_H-}--\ref{fig:ZBoson_gV-gA_H-} for a set
of models with various relations between $g_V$ and $g_A$. Similar to
the case of the hidden photon in Sec.\,\ref{sec:Photon}, the strongest sensitivity is exhibited to the model with mass of $Z'$ close to the threshold of nearly resonant production $e^+e^-\to Z'$. 

The signal dependence on the electron beam polarization may be exploited to reveal the nature of the hypothetical particle if the monophoton with missing energy events \eqref{missing} are observed. Indeed, flipping the beam polarization will change the signal rate unless $Z'$ is pure vector or pure axial. In all other cases it couples differently to left-handed and right-handed SM fermions, and so the signal rate changes with reversing the electron beam polarization. This reversing also changes the rate of neutrino production and hence increases the irreducible background \eqref{irr-back}. Naturally, the sensitivity of this mode is weaker, see below. However, it is worth noticing that the signal rate also depends on the polarization and relative sign of $g_A$ and $g_V$, and hence the signal can be larger for negative $\epsilon$ in a subset of models.

Generically, one can determine both $g_A$ and $g_V$ from the combined analysis of the signal events obtained with electron beam polarization of both signs. 

\begin{figure}[!htb]
    \centerline{
    \includegraphics[width=1.1\columnwidth]{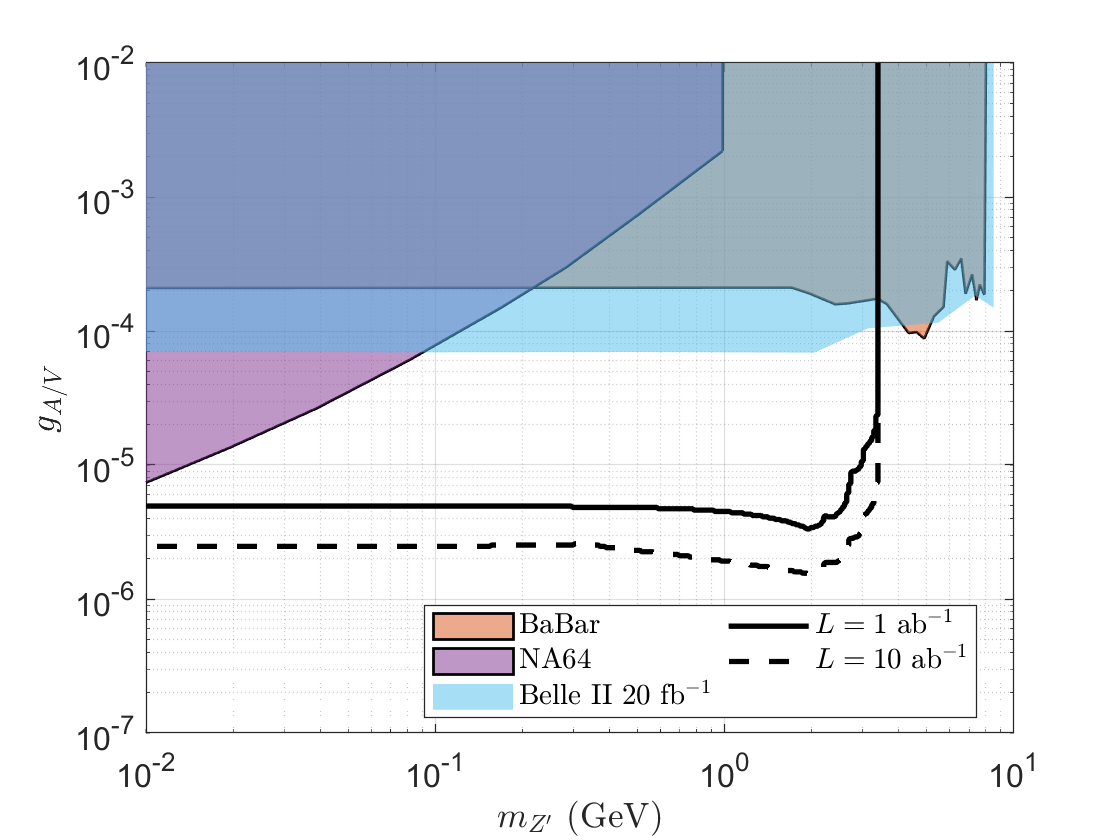}}
    \caption{The regions of the $Z'$-model with $g_V=g_A$ (the
      left-handed type coupling) to be probed atthe  $c$-$\tau$
      factory after one (solid black line) and ten (dashed black line)
      years of operation at 95\%\,C.L. with $e^+ e^- \rightarrow \gamma
      Z^{'}$ and $\epsilon=0.8$. We obtain the same one-year
      sensitivity to the models with $g_A=-g_V$ with SCTF operating
      with opposite beam polarization $\epsilon=-0.8$, while the
      ten-years sensitivity is worse by about 7\% in this case because
      of the higher background. Experimental limits (colored outlined
      regions) are the same as for the dark photon with substitution $\varepsilon^2e^2 = g_V^2+g_A^2$.}
    \label{fig:ZBoson_gV-gA_PH-_sum} 
\end{figure}

\begin{figure}[!htb]
    \centerline{
    \includegraphics[width=1.1\columnwidth]{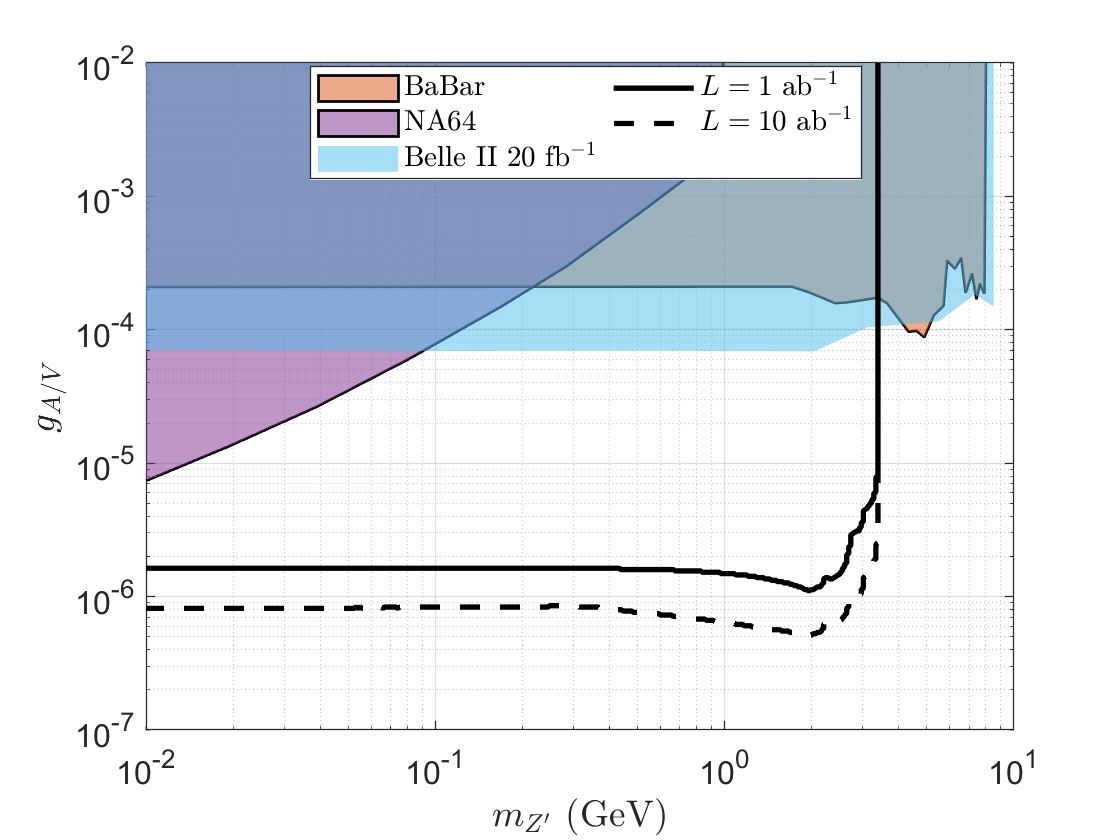}}
    \caption{The same as in Fig.\,\ref{fig:ZBoson_gV-gA_PH-_sum} but for the model with $g_V=-g_A$. The one-year sensitivity of SCTF with opposite beam polarization $\epsilon=-0.8$ to $Z'$ with $g_V=g_A$ are the same, while the ten-years sensitivity is by 7\% worse due to the bigger number of expected background events.}
    \label{fig:ZBoson_gV_NgA_PH-_sum}
\end{figure}

The best sensitivity is exhibited by the $c$-$\tau$ factory performing
the joint analysis of data collected over all the operation stages, as
we explained in Sec.\,\ref{sec:Background}. The constraints, expected
after one and ten operation years in accord with the plan depicted in Tab.\,\ref{Tab:schedule}, are outlined in Figs.\,\ref{fig:ZBoson_gV-gA_PH-_sum} and \ref{fig:ZBoson_gV_NgA_PH-_sum} for a set of models and positive and negative polarization modes. Recall that for purely vector and for purely axial $Z'$ coupling  the sensitivity coincides with that for the hidden photon outlined in Fig.\,\ref{fig:DarkPhoton_sum}. The signal rates are the same for models with opposite signs of the product $g_A\,g_V$ at SCTF operating with electron beams of opposite polarizations. However, the background rate is 1.5 times higher for the electron beam with negative longitudinal polarization, $\epsilon=-0.8$, and so the required number of signal events, according to \eqref{P}, is 1.15 times bigger at high statistics, when the numbers of expected background events exceed one. It gives a 7\% weaker sensitivity in this case (read the captions of Figs.\,\ref{fig:ZBoson_gV-gA_PH-_sum} and \ref{fig:ZBoson_gV_NgA_PH-_sum}).   

\begin{figure}[!htb]
    \centerline{
    \includegraphics[width=1.1\columnwidth]{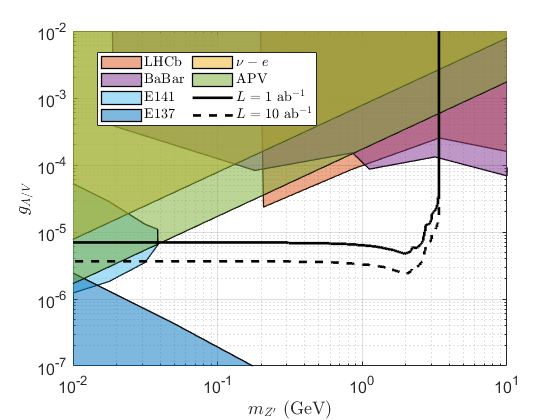}}
    \caption{The regions of the $Z'$-model for the HH case (50\% visible,
      50\% invisible modes) with $g_V=g_A$ (the left-handed type
      coupling) to be probed at the $c$-$\tau$ factory after one (solid black line) and ten (dashed black line) years of operation at 95\%\,C.L. with $e^+ e^- \rightarrow \gamma Z^{'}$ and $\epsilon=0.8$. We obtain the same one-year sensitivity to the models with $g_A=-g_V$ with SCTF operating with opposite beam polarization $\epsilon=-0.8$, while the ten-years sensitivity is worse by about 7\% in this case because of the higher background. Experimental limits (colored outlined regions) are taken from Ref.\cite{Cosme:2021baj}. }
    \label{fig:ZBoson_gV-gA_PH-_sum_vis}
\end{figure}

\begin{figure}[!htb]
    \centerline{
    \includegraphics[width=1.1\columnwidth]{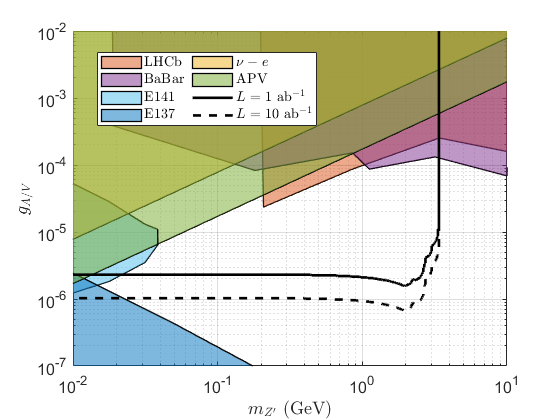}}
    \caption{The same as in Fig.\,\ref{fig:ZBoson_gV-gA_PH-_sum_vis} but for the model with $g_V=-g_A$. The one-year sensitivity of SCTF with opposite beam polarization $\epsilon=-0.8$ to $Z'$ with $g_V=g_A$ are the same, while the ten-years sensitivity is by 7\% worse due to the bigger number of expected background events.}
    \label{fig:ZBoson_gV_NgA_PH-_sum_vis}
\end{figure}

The searches at SCTF look very promising, and we further illustrate their power in Figs.\,\ref{fig:ZBoson_gV-gA_PH-_sum_vis} and \ref{fig:ZBoson_gV_NgA_PH-_sum_vis} 
with the HH model case introduced in Sec.\,\ref{sec:Photon}. 

At a $c$-$\tau$ factory operating with unpolarized lepton beams, the signal rate may be higher or lower as compared to that at SCTF, depending on the relative sign of the product $g_A\,g_V$ entering eq.\,\eqref{Z-cross}. The background is higher than that of SCTF with positive polarization $\epsilon=0.8$ and lower than that with negative polarization $\epsilon=-0.8$. Therefore, while at small statistics (when the searches are effectively background-free) the factory with unpolarized beams may exhibit higher sensitivity to models with a specific sign of $g_A\, g_V$ as compared to the polarized case, with growing statistics the SCTF becomes superior in testing the model with light $Z'$ for all values of the coupling constants. 

\section{Neutral (pseudo)scalar}
\label{sec:Axion}

While only one spin-zero fundamental particle, the SM Higgs boson, has been discovered so far, 
the nature may hide some other scalars and/or pseudoscalars. They can
originate from the extended SM Higgs sector, high energy grand
unification theory, scalar potential of a hidden sector, etc. They may
be naturally light, with the mass term protected from the quantum
corrections by symmetry arguments, e.g. like (pseudo-)Goldstone
bosons, the QCD axion providing the well-known example. Alternatively,
they may be light just because the entire hidden sector is
light. Within the portal paradigm, a scalar $s$, singlet with respect
to the SM gauge group, can couple to the visible sector via a so-called scalar portal: dimension-3 and dimension-4 renormalizable interaction terms with the SM weak Higgs doublet. At low energies this portal coupling induces the non-renormalizable interactions with photons, see e.g.\,\cite{Bezrukov:2009yw}
\begin{equation}
\label{scalar-photon}
    \mathcal{L} = \frac{1}{4} g_{s \gamma \gamma} \; s \, F_{\mu \nu} F^{\mu \nu}\,,  \end{equation}
where parameter $g_{s \gamma \gamma}^{-1}$ has dimension of mass. Similar coupling is predicted for light pseudoscalar $a$ or axion-like particles (ALPs) including the QCD axion, 
\begin{equation}
\label{axion-photon}
    \mathcal{L} = \frac{1}{4} g_{a \gamma \gamma} \; a \, F_{\mu \nu} \Tilde{F}^{\mu \nu}  
\end{equation}
where $\Tilde{F}^{\mu\nu}$ is the tensor dual to $F^{\mu\nu}$.  

In $e^+$-$e^-$ collisions the non-renormalizable couplings
\eqref{scalar-photon} and \eqref{axion-photon} produce via $s$-channel
photon exchange the (pseudo)scalar associated with a photon. If the
new particle is stable or decays predominantly into invisible
particles (e.g. from the hidden sector), it exhibits the signature
\eqref{missing} to be exploited in searches for these particles. As
far as these searches are concerned, there are no differences between the scalar and pseudoscalar particle cases, and below we perform the calculations in the ALP case, for concreteness. The same formulas are applicable in the scalar case as well.  

\begin{figure}[!htb]
    \centering
\begin{tikzpicture}
\begin{feynman}
    \vertex at (0,0) (i1);
    \vertex at (0,-2) (i2);
    \vertex at (1,-1) (a1);
    \vertex at (2,-1) (a2);
    \vertex at (3,0) (f1);
    \vertex at (3,-2) (f2);
    
    \diagram*{

    (i1) -- [fermion, edge label={\(e^{-}\)}] (a1),
    (i2) -- [anti fermion, edge label'={\(e^{+}\)}] (a1),
    (a1) -- [photon, edge label={\(\gamma\)}] (a2),
    (a2) -- [photon, edge label={\(\gamma\)}] (f1),
    (a2) -- [scalar, edge label'={\(a\)}] (f2),
    };
\end{feynman}
\end{tikzpicture}
    \caption{The Feynman diagrams for the production of axion and photon.}
    \label{diagram:axion}
\end{figure}
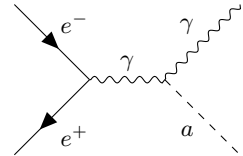

The process under study is $2\to 2$, see the diagram in Fig.\ref{diagram:axion}, 
so at a given angle $\theta_\gamma$ of the outgoing photon its energy $E_\gamma$ is fixed as \eqref{Egamma-2-2}. The corresponding differential cross section reads\,\cite{Tian:2022rsi}
\begin{equation}
\label{axion-cc}
    \frac{d\sigma}{d\cos\theta_\gamma} = 4\pi^2\alpha \frac{g^2_{a\gamma\gamma}}{(16\pi)^2} \left( 1 - \frac{m_a^2}{s} \right)^{\!\!3} (1+\cos^2\theta_\gamma)\,.
\end{equation}
One observes absence of any amplification of \eqref{axion-cc} in the
limit $m^2\to s$ contrary to the vector \eqref{cross-A} and
pseudo-vector \eqref{Z-cross} cases: the resonant (pseudo)scalar
production, $e^+e^-\to a$ cannot proceed through the $s$-channel
exchange of the spin-1 particle, photon. There is no dependence on the electron polarization, similar to other processes considered in Secs.\,\ref{sec:Photon} and \ref{sec:MCP}. 

\begin{figure}[!htb]
    \centerline{
    \includegraphics[width=1.1\columnwidth]{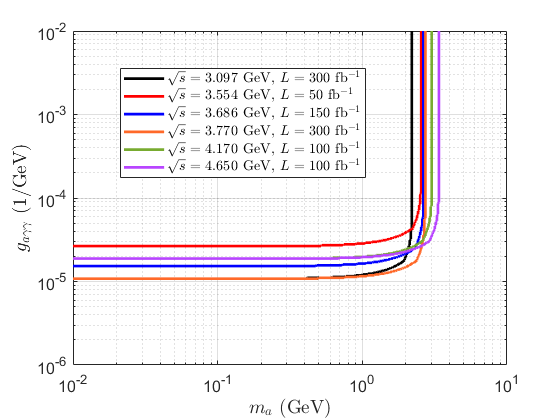}}
    \caption{The regions above the lines will be tested at 95\%\,C.L. with $e^+ e^- \rightarrow \gamma a$.}
    \label{fig:Axion}
\end{figure}

We integrate eq.\,\eqref{axion-cc} over the photon angle $\theta_\gamma$ within the adopted cuts \eqref{photon-cut} and requiring 3 events obtain the expected sensitivity at 95\% C.L. The results are presented in Fig.\,\ref{fig:Axion}.  
The limiting lines are flat at small masses, and they steadily grow up with the mass approaching the reaction threshold. 

\begin{figure}[!htb]
    \centerline{
    \includegraphics[width=1.1\columnwidth]{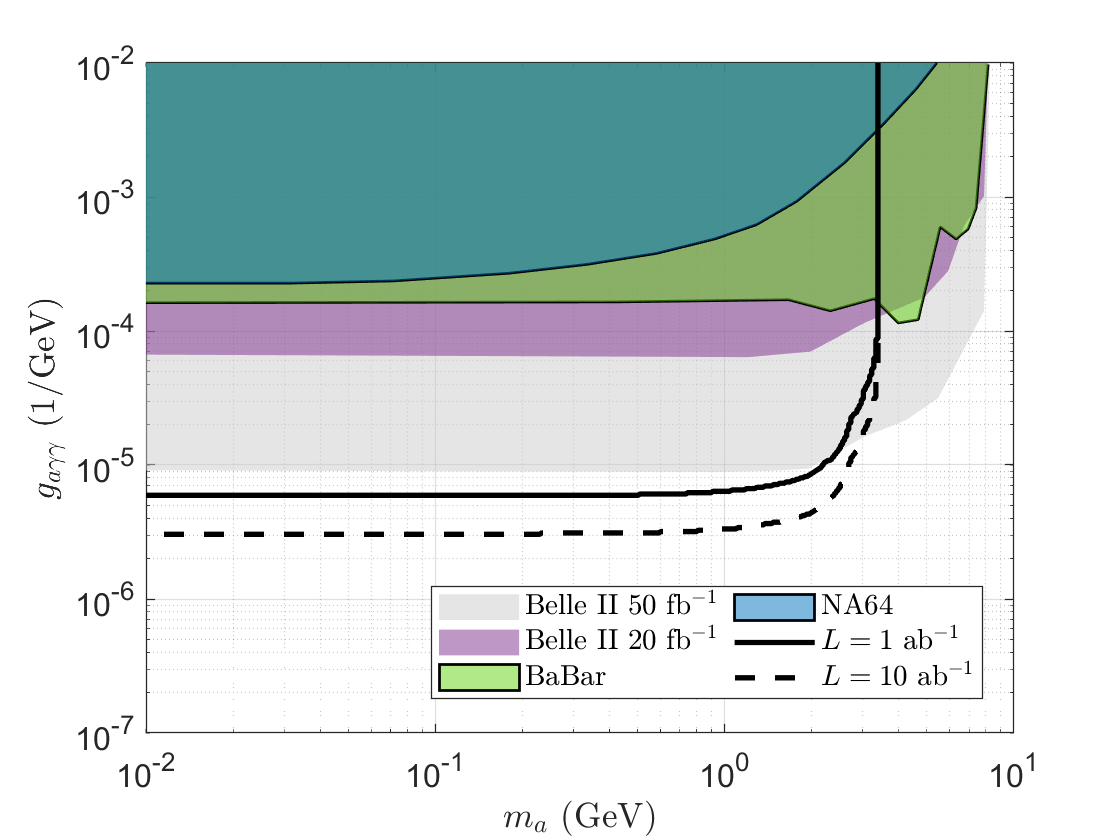}}
    \caption{Bounds on the axion coupling constant to photons $g_{a\gamma\gamma}$. The regions above the black bold and dashed lines will be explored  at 95\%\,C.L. with $e^+ e^- \rightarrow \gamma a$ after one and ten years of operation, correspondingly. The existing limits (colored and outlined) and expected reaches of ongoing experiments (colored) are taken from Ref.\,\cite{Darm__2021}.}
    \label{fig:Axion_sum}
\end{figure}

\begin{figure}[!htb]
    \centerline{
    \includegraphics[width=1.1\columnwidth]{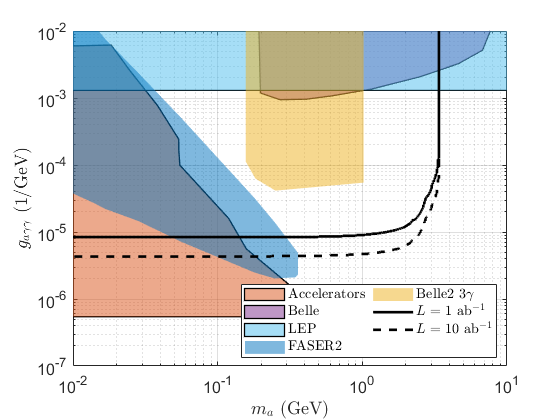}}
    \caption{Bounds on the axion coupling constant to photons
      $g_{a\gamma\gamma}$ in the HH model. The regions above the black bold and dashed lines will be explored  at 95\%\,C.L. with $e^+ e^- \rightarrow \gamma a$ after one and ten years of operation, correspondingly. The existing limits (colored and outlined) and expected reaches of FASER, Belle2 (3$\gamma$) and Accelerator experiments (colored) are taken from Ref.\,\cite{Feng:2022inv}. Limits from Belle and LEP are taken from Ref.\,\cite{PhysRevLett.125.161806}. }
    \label{fig:Axion_sum_vis}
\end{figure}

The data, collected over one and ten years of operation allows one to place stronger constraints on the light axion coupling to photons. Performing the calculations outlined in Sec.\,\ref{sec:Background}, we obtain the expected limits presented in Fig.\,\ref{fig:Axion_sum} 
(see Fig.\,\ref{fig:Axion_sum_vis} for HH case) 
along with current bounds and constraints expected from the ongoing experiments. 

The same sensitivity as in Figs.\,\ref{fig:Axion} and \ref{fig:Axion_sum} will be exhibited by SCTF to coupling \eqref{scalar-photon} of a light singlet scalar and photons. 

Note in passing that since the signal differential cross section
\eqref{axion-cc} does not depend on the beam polarization, the signal
rate must be the same for a $c$-$\tau$ factory with unpolarized
beams. The usage of polarized beams enables one to additionally
suppress the background. Therefore, similar to the case of a hidden photon described in Sec.\,\ref{sec:Photon}, for the unpolarized beams we would obtain somewhat lower sensitivity to the axion coupling, especially for the joint analysis of ten-years statistics. 

\section{Discussion}
\label{sec:Discussion}

To summarize, we estimate the sensitivity of the proposed super $c$-$\tau$ factory 
SCTF \cite{Charm-TauFactory:2013cnj,Bondar:2019zgm,Epifanov:2020elk} to physical parameters of a set of models with hypothetical light particles whose production in $e^+$-$e^-$ collisions can be associated  with a single photon and missing energy events. The electron beam in the project of SCTF will be polarized, which mitigates the background and gives better chances to probe the models with smaller couplings as compared to the factory exploiting unpolarized beams. We find the SCTF can explore models with one-two orders of magnitude smaller couplings that the previous experiments. Depending on the model and parts of the model parameter space the SCFT prospects in searches for the new particles are either complementary to those of the project under construction or more promising. Note, there are other signatures which may be promising for testing at SCTF particular models with light feebly interacting particles. Our study reveals that even in models with presence of noticeable visible decay modes of the light exotic particles, searches for the invisible channels at SCTF are still competitive even with the ongoing searches for the visible modes.   

The reversibility of the electron beam polarization can be effectively used to reveal the nature of the hypothetical particles. In most cases the sensitivity to the new particle couplings are almost mass-independent for sufficiently light particles. If the signal is observed, the mass of new particles can be measured for the moderate range of masses with accurate determination of the photon energy. However, the resolution of the electromagnetic calorimeter is always finite, and for light particles it will allow one to place only an upper limit on mass of the hypothetical particle possibly responsible for the signal events. A specific investigation can yield a lower bound on the mass of some hypothetical candidates (or discover its nature and measure its mass): the very light millicharged particles start to interact inside the detector material, see e.g.\,\cite{Arefyeva:2022eba}, light hidden photons oscillate to visible photons, 
see e.g.\,\cite{Demidov:2018odn}, etc.   

It is worth mentioning that while we present our numerical estimates for a particular project of the super $c$-$\tau$ factory developing in BINP (Novosibirsk), the similar study can be performed for the project of super $\tau$-$c$ factory in China. A brief look through the literature shows that at the present stage of development of both projects the main ingredients essential to the present study (beam polarization, beam energy range, calorimeter energy resolution, etc) are very similar, and hence similar are chances to observe the new exotic particles.

\section*{Acknowledgements}

We thank A.\,Bondar, S.\,Demidov, S.\,Gninenko, D.\,Kirpichnikov, I.\,Logashenko and P.\,Pakhlov for the valuable discussions and L.\,Darme for correspondence.  The work on the signal evaluations is supported by the Russian Science Foundation RSF grant 21-12-00379. The work of D.\,K. on the background estimates is supported by the Foundation for the Advancement of Theoretical Physics and Mathematics “BASIS”. 

\bibliography{ref}
\end{document}